\documentclass[twocolumn,secnumarabic,amssymb, nobibnotes, aps, prd]{revtex4-2}

\usepackage{graphicx,float,amsmath}
\usepackage{mathrsfs}
\usepackage{natbib}

\begin{document}

\date{\today}

\title{Anomalous ambipolar  transport in depleted  GaAs nanowires}

\author{H. Hijazi$^1$}
\author{D.~Paget$^2$}
\author{A C. H. ~Rowe$^2$}
\author{G. Monier$^1$}
\author{K. Lahlil $^2$}
\author{E. Gil$^1$}
\author{A. Trassoudaine$^1$}
\author{F.~Cadiz$^2$}
\author{Y. Andr\'e$^1$}
\author{C. Robert-Goumet$^1$}

\affiliation{%
$^1$ Universit\'e Clermont Auvergne, Clermont Auvergne INP, CNRS, Institut Pascal, F-63000 Clermont-Ferrand, France}

\affiliation{%
$^2$Physique de la mati\`ere condens\'ee, Ecole Polytechnique, CNRS, Universit\'e Paris Saclay, 91128 Palaiseau, France}

\begin{abstract}
We have used a polarized microluminescence technique to investigate photocarrier  charge and spin transport in n-type  depleted GaAs nanowires ($ \approx  10^{17}$ cm$^{-3}$ doping level). At 6K, a long-distance tail appears in the luminescence spatial profile, indicative of  charge and spin transport, only limited by the length of the NW. This tail is independent on excitation power and temperature. Using  a self-consistent calculation based on the  drift-diffusion and Poisson equations as well as on photocarrier statistics (Van Roosbroeck model), it is found that this tail  is due to photocarrier drift in an internal  electric field nearly two orders of magnitude larger than  electric  fields predicted by the  usual ambipolar model. This large electric field appears because of two effects. Firstly,  for transport in  the spatial fluctuations of the conduction band minimum and valence band maximum,  the electron mobility is activated by the  internal electric field. This implies, in a counter intuitive way,  that the spatial fluctuations favor long distance transport.   Secondly, the range of carrier transport  is further increased  because of the finite NW length, an effect which plays a key role in one-dimensional systems.
\end{abstract}
\maketitle

\section{Introduction}
\label{intro}
In the past few years,  investigations of transport in semiconductor nanowires (NW's) have gained interest because of  potential applications to solar cells \cite{krogstrup2013}, lasers  \cite{duan2003} and quantum computing  \cite{vandenberg2013}.  For GaAs, it has been reported recently that   GaAs NW's grown on Si substrates have strong potentialities  for charge and spin transport \cite{hijazi2021}. These NW's are n-doped in the low $ 10^{17}$ cm$^{-3}$ range, and are therefore on the metallic side of the Mott transition \cite{benzaquen1987}.   They are well-adapted for spin transport, since the donor concentration nearly corresponds with that of the maximum of the spin relaxation time \cite{dzhioev2002}, thus ensuring conservation of spin polarization over large distances.\

It may be thought that, in such NW's,  transport of photocarriers should be  difficult because of the presence of spatial fluctuations of the energy of the top of the valence band, induced by statistical spatial fluctuations of the donor concentration \cite{efros1972, shklovskii1984}. However, it has been shown that  carrier transport in this disordered system can occur over distances as large as  25 $\mu$m \cite{hijazi2021}. Several phases in the spatial profiles have been observed, due to i)  the buildup of internal  electric fields   which modify the photocarrier mobilities and ii)  to  the  subsequent spatial redistribution of the Fermi sea for undepleted NW's. However, no    interpretation for these results has been proposed.  \

The present work is an experimental and theoretical analysis of  charge and spin transport in   NW's grown on Si substrates.  We have chosen depleted NW  so that the charge spatial profiles  merely reveal the buildup of the internal electric field since there is no Fermi sea.  The spatial charge profile   exhibits  a relatively fast decrease followed by a slow tail, which  weakly depends on   excitation power and temperature.   As found by numerical resolution of conservation equations, this tail is caused by drift transport in  an  internal electric field $E$ of a fraction of a V/$\mu$m. These results are at  variance with the predictions of  the usual ambipolar model  \cite{smith1978, zhao2009, cadiz2015c, cadiz2015d, cadiz2017b, paget2012}  which  predicts internal electric fields smaller by two orders of magnitude. Such large internal field is shown to build up for two reasons. Firstly, as expected for such doping level,  the mobility of photoelectrons  depends on the internal electric field \cite{baranovskii2006}. This implies that, for charge and spin transport, metallic NW's  appear as better candidates than NW's on the insulating side of the insulator/metal transition.  Secondly,  the  electric field is further amplified by the finite size of the NW.  It is anticipated  that such large electric fields are specific to one-dimensional systems.\

\section{Experimental}  

\subsection{Principles }

Here we study   NW's HVPE-grown on Si(111) substrates using gold-catalysis  at 715  $^\circ$C \cite{hijazi2018}. In order to reduce the surface recombination velocity, the NW's were chemically treated by  a low alkaline (pH $\approx$ 8.5) hydrazine sulfide solution. This produced a negligible NW etching by the solution and covered the surface by a nitride layer so that surface recombination was equivalent to that of the nearly ideal Ga$_{1-x}$Al$_{x}$As/GaAs interface \cite{alekseev2015}. After passivation, the NW's, standing on the substrate, were scraped and deposited horizontally on a grid of lattice spacing 15 $\mu$m. The results presented here  were obtained on a  depleted NW  of  length 20  $\mu$m and  of diameter 100 nm that is smaller than the limit of  180 nm for NW depletion  \cite{hijazi2021}.\

The NW was excited at 6K by  a tightly-focused, continuous-wave, laser beam (Gaussian radius $\sigma \approx 0.6\; \mu$m, energy $1.59$ eV). Spatially-resolved spectral analysis of the intensity and circular polarization of the luminescence was performed using a setup described elsewhere  \cite{favorskiy2010, hijazi2021}.  Using  liquid crystal modulators, the sample was excited  with $\sigma^{\pm}$-polarized light and the intensity $I(\sigma ^{\pm})$ of the luminescence components with $ \sigma ^{\pm}$ helicity was selectively monitored.  The  luminescence intensity is the sum of these two components and given by  
 
\begin{equation}\label{tau}
\mathscr{I} =  K (n+n_0)p 
\end{equation} 
where $n$ is the  photoelectron concentration, $p$ is the hole concentration and $K$ is the bimolecular recombination coefficient. Here,  quite generally, we take a nonzero  electron concentration in the dark $n_0$. This value will be  zero for the experimental depleted NW's but will have a weak  nonzero  value for computations.  The difference signal $ \mathscr{I}_{D}= \mathscr{I}(\sigma ^+) - \mathscr{I}(\sigma ^-)$ is equal to $Kp \mathscr{P}_i  s$, where $ \mathscr{P}_i= \mp 0.5$ for $ \sigma ^{\pm}$- polarized excitation. Here $s=n_+ - n_-$, where $n_{\pm} $  are the concentrations of electrons with spin $\pm 1/2$, choosing the  excitation light direction  as the quantization axis, is the spin density. \ 

\begin{figure}[tbp]
\includegraphics[clip,width=9 cm] {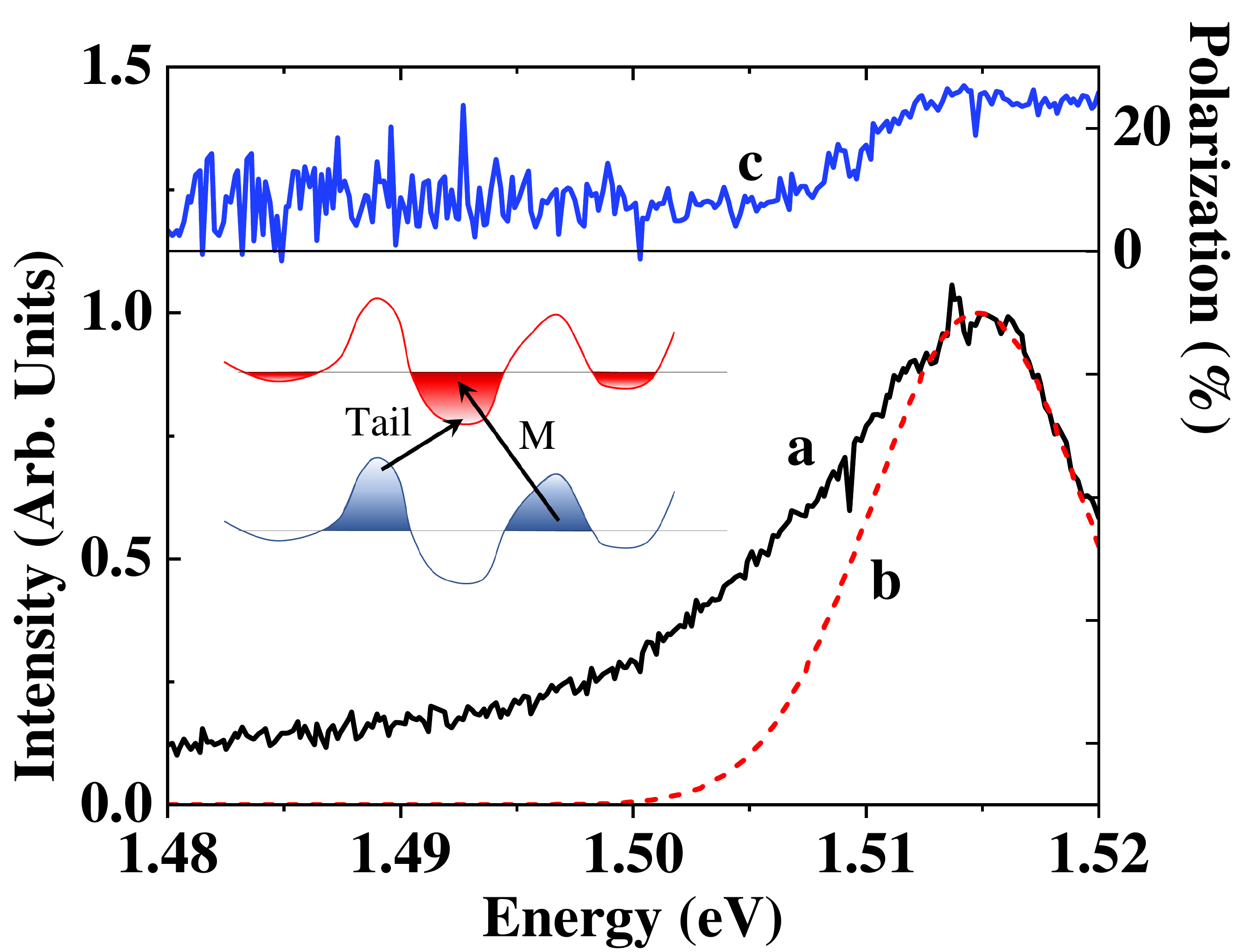}
\caption{Curve a shows the NW intensity spectrum at the excitation spot  for a small excitation power of 45 $\mu$W.  Also shown is a Gaussian fit of the main line (Curve b) which reveals a  low-energy tail which extends down to $1.48$ $eV$. Curve c shows the corresponding polarization spectrum and reveals a large photoelectron spin polarization of nearly $50 \%$. Shown in the inset is an illustration of the spatial fluctuations of the conduction and valence band, illustrating the mechanisms for recombination of the main line (M)  and of the tail.}
\label{Fig01}
\end{figure} 

\subsection{Results}
The nearbandgap luminescence  and polarization spectra, taken at the excitation spot,  are shown in Fig. \ref{Fig01}. The lattice temperature is 6K and a very small  excitation power of  $45$ $\mu$W is chosen.  More details are shown in the  supplementary material. The luminescence spectrum  peaks near the bandgap energy. It  can be decomposed into a main line, approximated by a gaussian lineshape of half-width 6.5 meV  (curve b), and  a low-energy tail which extends down to 1.48 eV. As shown in the inset of  Fig. \ref{Fig01}, this tail is attributed to spatially indirect transitions, where the transition energy is lowered  by local electric fields in the fluctuations. \
 
Curve c of   Fig. \ref{Fig01} shows the  corresponding polarization  spectrum. For energies larger than 1.51 eV, the polarization has a  very large value above 20 $\%$, implying a photoelectron spin polarization close to the maximum value of $ |\mathscr{P}_i| =  50  \%$. This suggests a very large spin relaxation time, as predicted for this sample where the Bir-Aronov-Pikus (BAP) process is weak because of the weak hole concentration \cite{bir1975, dzhioev2002}.  Note that the polarization  increases near 1.504 eV, which coincides with the onset of the gaussian component of the intensity spectrum. This is because the  electron quasi Fermi level lies above the minimum of the conduction band fluctuations so that  electrons below this level are degenerate. \

\begin{figure}[tbp]
\includegraphics[clip,width=8 cm] {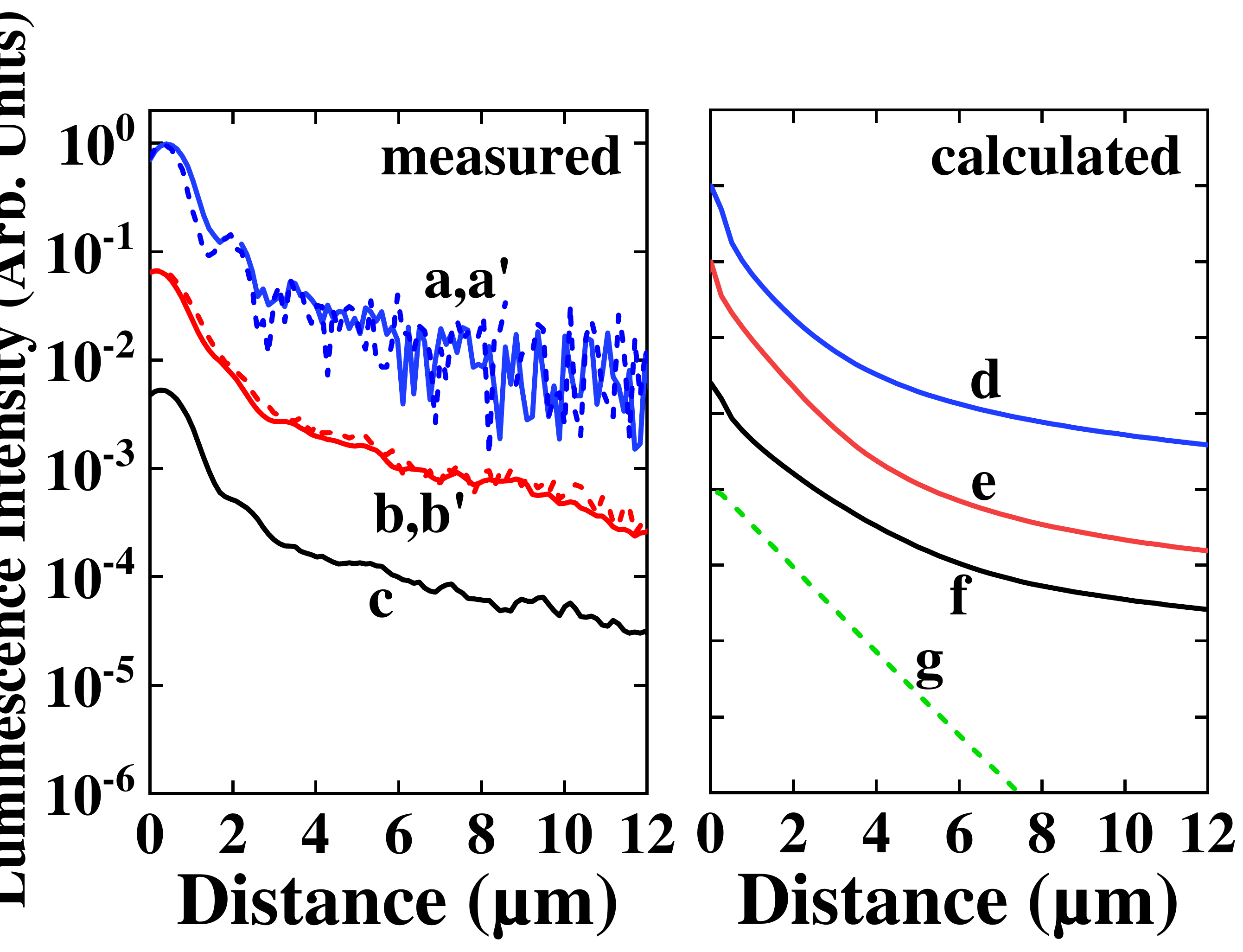}
\caption{ Curves a and a'   show the intensity (solid line) and difference (dotted line)  spatial profiles  at 6K and for an energy of 1.512 eV, for an excitation power of 45 $\mu$W. Shown in curves b  and b' are the corresponding results for an excitation power of 1 mW.  Curve c shows the intensity spatial profile at a temperature of 30K. The right panel shows corresponding profiles calculated using the model of Sec. III, corresponding respectively to an excitation power of  45 $\mu$W (Curve d), 1  $m$W (Curve e), a temperature of 60K(Curve f). Curve g shows the profile calculated for an excitation power of  45 $\mu$W  (same as Curve d), but without a dependence of the electron mobility on electric field ($E_e=0$)}
\label{Fig02}
\end{figure}

Shown in curve a of  Fig. \ref{Fig02} is the intensity  spatial profile, taken in the same conditions as curve a of  Fig. \ref{Fig01}.  Curve a' shows the profile of the difference signal  $ \mathscr{I}_{D}$. The two spatial profiles are quite similar. This shows that  the photoelectron spin polarization  $ \mathscr{I}_{D}/( \mathscr{I}  \mathscr{P}_{i})$ is constant over the profile. This  is expected in our case where the spin relaxation time is large, as suggested by  the large photoelectron spin polarization. The spatial profile  is composed of  a rapid decrease up to 2 $\mu$m, with a  slower decrease  for larger distances, superimposed on fluctuations which are reproducible from one curve to the other. These fluctuations may be caused by inhomogeneities of the surface recombination or by uncomplete  spatial averaging of the microscopic potential fluctuations described in the inset of Fig. \ref{Fig01}.  \

Such shape is very different from  a single exponential, which is  the predicted profile  for one dimensional  unipolar transport (see supplementary information of Ref. \cite{hijazi2021}).   The observation  of the long distance tail implies that, although  most  carriers recombine near the place  of excitation,  a significant fraction escape from the potential fluctuations at the place of excitation and can  be transported  over large distances. Since hopping processes are easier for electrons than for holes because of their weaker effective mass,  there builds up an outwards  internal electric field which in turn drives photoholes  out of the excitation spot  provided its magnitude  is comparable   to that of the electric field of the fluctuations (of the order of the unscreened effective field near a donor  $E_D/a_0^* \approx 0.6$ $V/\mu$m, where $E_D$ is the donor binding energy and $a_0^* $ is the effective Bohr radius).  \ 

Shown in curves b  and b' of  Fig. \ref{Fig02} are the intensity and difference spatial profiles obtained  for a larger excitation power of 1 mW. In the same way as for a smaller excitation power, the two curves are similar, revealing that the photoelectron spin polarization does not decrease during transport. Comparison between Curves a and b  reveals that, unlike observed earlier  for ambipolar transport in 3D samples \cite{smith1978, zhao2009, cadiz2015c, cadiz2015d, cadiz2017b, paget2012},  the excitation power has little  effect on the spatial profiles. \

Curve c was taken in the same conditions as Curve b  but  at a higher lattice temperature of 30K, leading to a  decrease of luminescence intensity by about one order of magnitude. This curve is quite similar to Curve a and Curve b, implying that temperature has a weak effect on the spatial profile. Such result may appear surprising, in view of the strong temperature dependence of the conductivity    reported for metallic systems \cite{benzaquen1987}.\

\section{ Interpretation}

The  results of the preceding section show that  depleted NW's on the metallic side of the insulator/metal transition appear as ideal candidates for charge and spin transport. Transport exhibits a long distance tail up to 15 $\mu$m, mostly  limited by the NW end. The photoelectron polarization at the excitation spot  is close to its maximum value determined by the transition probabilities   and weakly  decreases during transport.  \

 In order to interpret these results,  calculation of  the spatial distributions of electrons and holes was performed using  the Van Roosbroeck model  \cite{vanroosbroeck1953}.  For holes, the drift-diffusion equation  is 

\begin{equation}\label{eqdiftrous}
	 g  - \mathscr{I} -  \frac{p}{\tau _{nr}^h} -\vec{\nabla } \cdot  [\vec{J }_p/q ]= 0. 
\end{equation} 

Here $g$ is the rate of creation of electron-hole pairs, $q$ is the absolute value of the electron charge,  $\tau _{nr}^h$ is the hole nonradiative recombination time and $\vec{J}_p$ is the hole current. The  corresponding spin-unresolved equation for electrons is 

\begin{align}
\label{eqdifel}
	 g  - \mathscr{I} -  \frac{n}{\tau _{nr}^{e}} +  \vec{\nabla } \cdot [\vec{J }_n/q ]= 0.  	  
\end{align}
 	
where $\tau _{nr}^e$ is the electron nonradiative recombination time. The quadratic dependence of the spatially-integrated luminescence  intensity on excitation power (see Supplementary Material) shows that nonradiative recombination is dominant over radiative recombination. Since the two recombination terms must be equal after spatial integration which removes the effect of transport, and within the  hypothesis of  charge neutrality (i. e. that the total photoelectron and photohole charges are equal) one can assume  that $\tau _{nr}^{e} = \tau _{nr}^{h} = \tau $. The electron and hole currrents are given by 
  
\begin{equation}    
\vec{J}_n = q\mu_n (n+n_0) \vec{\nabla }E_{Fn} = q \mu_e (n +n_0) \vec{E} + q D_e \vec{\nabla }n 
 \label{Jn} 
\end{equation} 
 
and  
 
\begin{equation}    
\vec{J}_p = q\mu_p p \vec{\nabla }E_{Fp}= q \mu_e p \vec{E} - q D_h \vec{\nabla }p 
 \label{Jp} 
\end{equation} 

where $\mu_{e(h)}$ are the electron (hole) mobilities and  $D_{e(h)}$ are the corresponding diffusion constants. Here   $E_{Fn}$  ( $E_{Fh}$) is the energy of the electron (hole) Fermi level  with respect to its value at equilibrium. The electronic concentration can be expressed by Boltzmann statistics
\begin{equation}    
n= N_c  \exp   \frac{E_{Fn} -qV-E_c }{ k_BT_e}  
 \label{n} 
\end{equation}
where   $k_B$ is Boltzmann's constant, $T_e $ is the photocarrier temperature and $E_c$ is the energy of the bottom of the conduction band.  As discussed in the supplementary material, the hole energy distribution for a depleted NW is closer to a Boltzmann one than for an undepleted one. This distribution will be approximated by 

\begin{equation}    
p= N_v  \exp   \frac{-E_{Fp} +qV+E_v}{ k_BT_e}  
 \label{p} 
\end{equation}
 
 where $E_v$ is the energy of the top of the valence band.   Here $N_c $ ($N_v $)  is the effective density of states of the conduction (valence) band, and   $E_c$ is the energy of the bottom of the conduction band. Here $V$ is  the spatially-dependent potential, given by  Poisson's equation, which can be written, for a spatially homogeneous doping
 
\begin{equation}    
\epsilon _s  \frac{d^2V}{dz^2}= q(N_d +p-n-n_0) 
 \label{Poisson} 
\end{equation}

where  $\epsilon _s $ is the static permittivity.   
 
\begin{figure}[tbp]
\includegraphics[clip,width=9 cm] {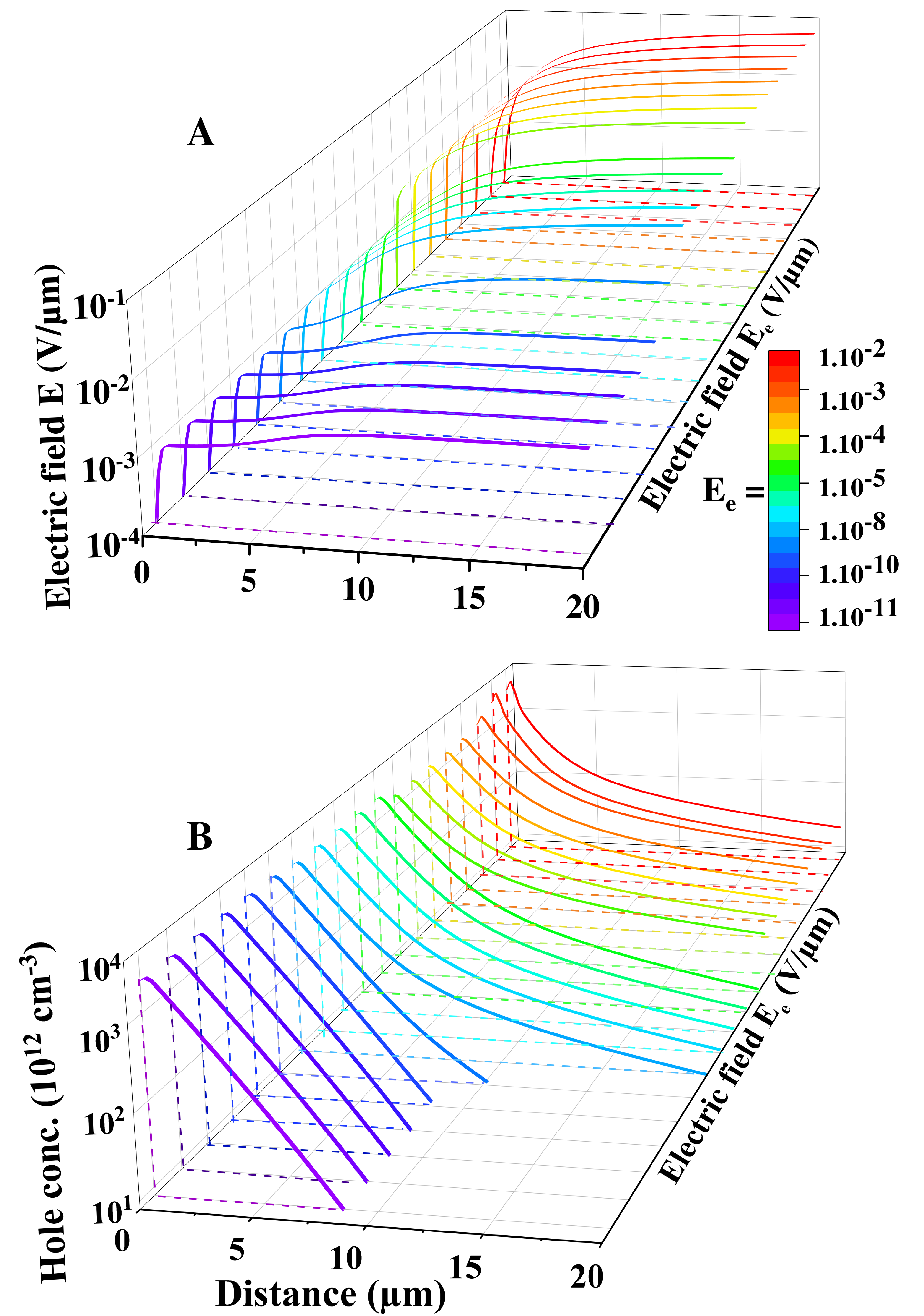}
\caption{Panel A shows the calculated spatial profiles of the hole concentration for increasing  values of $E_e$.  Panel B shows the corresponding spatial profiles of the internal electric field.  For a very small value of $E_e$ the decay is rapid, corresponding  to the usual ambipolar case as described by Eq. \ref{champel1}. Upon increase of  $E_e$, a tail appears in the charge profile, while a large internal electric field builds up at large distance.  Panel C  shows the spatial profiles of  electron diffusive (Curve b) and drift (Curve c) currents, as well as of the hole diffusive ( Curve d ) and drift (Curve c) currents, calculated with $E_e= 10^{-3}$ V/$\mu$m.  Panel D shows the electric field value at  5 $\mu$m from the excitation spot, as a function of the NW length. The strong reduction of electric field for a length larger than 170 $\mu$m demonstrates the length-dependent transport. }
\label{Fig03}
\end{figure}

For   NW's on the metallic side of  the insulator/metal transition,  transport occurs through hopping processes  assisted by the electric field. This results in a dependence of the  mobility on the electric field \cite{baranovskii2006, cleve1995, hijazi2020b}, given by\

\begin{equation}    
\mu_{e }(E)= \mu_{e}^* \exp\left[-\left(\frac{E_{e}}{\sqrt{E^2+ E_T^2}}\right)   \right] \approx  \mu_{e}^*  \exp\left[-\left(\frac{E_{e}}{|E|}\right)   \right]  ,   
 \label{mob} 
\end{equation}

where $\mu_{e}^*$ is  the mobility at large electric fields. The electric field $E_{e}$ is given by

\begin{equation}\label{Ee}
	  E_{e} =\frac{\Delta_{e}}{q \delta}  , 
\end{equation}  \ 
 
where $\Delta_{e}$ is a  characteristic energy,  $\delta$ is the length of an elementary hopping process,   and  $E_T = k_BT_e/(q \delta)$. Here, one will use  the approximate expression in  Eq. \ref{mob} since, as shown by the weak effect of temperature on the profile, one probably has  $E_e>> E_T$.  One may think that the hole mobility  also  depends on electric field.  However,  such dependence has no effect on the spatial profiles, since hole  diffusive and drift currents are negligible with respect to their  electronic  counterparts and will not be included here   \cite{note13}. \
	
The coupled equations   \ref{eqdiftrous},  \ref{eqdifel} ,  \ref{n},  \ref{p} and \ref{Poisson} must be solved self-consistently.  The calculations give  the spatial distribution of $V$,   $E_{Fp}$  and  $E_{Fn}$ and subsequently  the spatial profiles of $n$, $p$ and $E$.  Finally, the spatial profile of the luminescence intensity is given by  Eq. \ref{tau}.\

 Independently, the expression of the internal electric field  is obtained by comparing  Eq. \ref{eqdiftrous} and  Eq. \ref{eqdifel}. This gives \

\begin{equation}\label{cour}
	 \vec{\nabla } \cdot [\vec{J }_e  +  \vec{J }_h] = \frac{q(n-p)}{\tau}  
\end{equation}  \ 

where  $ \vec{J_e}$ and $ \vec{J_h}$, given by Eq. \ref{Jn}  and  Eq. \ref{Jp}, respectively,  depend on electric field via the contribution of drift currents.  Within what will be called below  the usual ambipolar model, one assumes a spatially-infinite sample with negligible   charge concentrations and electric field at its end \cite{smith1978,   zhao2009,  cadiz2015c}. One also assumes  charge neutrality ($n=p$). Thus, the sum of electron and hole currents  which is zero at infinity,  is also zero at all points in the NW, so that  

\begin{equation}\label{champel1}
	\vec{E}  \left[ \mu_e (n+n_0) +\mu_h p \right]  =  D_h  \vec{\nabla} p -  D_e \vec{\nabla}n  
\end{equation} 
 
Although,  as seen in Sec. IV below,  Eq. \ref{champel1}  must be modified to correctly interpret the long distance tail,   it allows us to understand  the weak dependence of the luminescence intensity profiles on excitation power. Indeed, provided $n>>n_0$, multiplication of electron and hole concentrations by a common factor will not affect the electric field and therefore the shape of the spatial profile. The  weak dependence of the spatial profile on temperature is in agreement with the observed weak temperature dependence of the  photoconductivity of disordered systems  \cite{baranovskii2006, cleve1995}. Such effect can  be understood if, in Eq. \ref{mob},  $E_T << E$.  In this case, the electron mobility nonlinearly depends on electric field according to the approximate Eq. \ref{mob} and weakly depends on temperature. \

For solving the above coupled equations, we applied a Newton-Raphson algorithm, as described in Ref. \cite{farrell2017}, to  a NW of length $20$ $\mu$m with an excitation spot at $5$ $\mu$m  from the end.  For the boundary conditions, one imposed a zero potential at the NW ends and a zero recombination current at the lateral surfaces. In order to avoid divergence,  a nonzero value of $n_0$  and a zero value of  $E_e$ were used as a starting point for the calculations. The quantity $n_0$ was  subsequently decreased to   $5\times 10^{13}$  cm$^{-3}$ and $E_e$  was progressively increased to above $ 10^{-2}$ V/$\mu$m.  The high-field mobility values were   $\mu_{e}^*= 10^4$  cm$^2/Vs$ and    $\mu_{h}^*= 3 \times 10^3$  cm$^2/Vs$ that is,  slightly larger  than  mobilities  for a degenerate doping level \cite{lovejoy1995}. This is probably because the absence of intrinsic electrons increases the electron and hole collision time.  The  values of the other parameters  were found to have  a negligible effect on the profile. The electron and hole lifetimes were $\tau =1$ns. The temperature $T_e$   was taken to 30 K. \

We first  consider the case of  a doping level slightly smaller than  the insulator/metal transition. Since the spatial fluctuations of the conduction band minimum and valence band maximum are negligible \cite{lowney1986}, one can take  $E_e =0$.  The   calculated  luminescence spatial profile is shown in Curve  g of  Fig. \ref{Fig02}.  The  excitation power was $\approx 20$ $\mu$W i. e. close to that used in  Curve a.  The profile exhibits a rapid,  approximately exponential, decrease, nearly independent on excitation power and does not interpret the experimental results. At the excitation spot, the electric field is zero for symmetry reasons. Away from the excitation spot, the electric field is  estimated  using Eq. \ref{champel1},   to  $\approx  D_e/(\mu_e  L_d) $  where $L_d $  is the exponential slope of the decrease. Using $L_d \approx 0.8 \mu$m and the Einstein relation $D_e = \mathscr{E} \mu_e/q$, where   $\mathscr{E}$ is an energy of the order of the band fluctuation amplitude  \cite{baranovskii2006} and assuming charge neutrality, we obtain   $E  =3 \times  10^{-3}$ V/$\mu$m. This relatively weak value implies that diffusive currents are larger than drift currents and  explains the absence of a long-distance tail. It is concluded that NW's of a doping level on the insulating side of the Mott transition  are not  expected  to be good candidates for charge and spin transport. \

For increasing  values of $E_e$,  the calculated photohole concentration profiles for an excitation power of  150 $\mu$W are shown in Panel A of  Fig. \ref{Fig03}.  Panel B shows the corresponding spatial profiles of the electric field. In agreement with the experimental results, and provided  $E_e >    10^{-5}$ V/$\mu$m, the calculated concentration profiles exhibit  a long distance tail at a distance larger than $5$ $\mu$m . In this regime,   the profile weakly depends on $E_e$.  An estimate  of  $E_e $ can be  obtained, using  Eq. \ref{Ee}  taking  $\Delta_{e} = 1$  meV, corresponding to the amplitude of the fluctuations  \cite{lowney1986} and  $\delta_{e}= 50$ $nm$. One  finds $E_e= 0.2 $ V/$\mu$m. This is a high estimate of $E_e$ since the electron  hopping process  may occur over larger distances. However,  $E_e$  is in all cases larger than $10^{-5}$ V/$\mu$m so that a long distance tail should appear.\

The internal electric field  at a distance from the excitation spot larger than $2$ $\mu$m is of   several  $ 10^{-2}$ V/$\mu$m i. e. nearly two orders of magnitude larger than for the ambipolar case.  This suggests  that the  tail in the hole concentration profile is caused  by outward hole and electron drift in the electric field. \

In order to confirm this hypothesis, we have  calculated   the spatial profiles of the electron and hole currents, using  $E_e= 10^{-3}$ V/$\mu$m. As seen in Fig. \ref{Fig04}, two spatial phases are visible. Up to a distance of $1.5$ $\mu$m, because of the large concentration gradient, diffusive currents are larger than drift currents. For larger distances, drift currents indeed predominate, because of the large electric field.  These currents explain the  presence of charge and spin transport over record distances. Note that the electron drift (Curve a) and  diffusive (Curve b)  currents are, as expected, dominant over their hole counterparts (Curves c and d respectively). This  justifies the hypothesis taken above of a negligible  field dependence of the hole mobility \cite{note13}.\  

Curve d of Fig. \ref{Fig02} shows  the calculated intensity spatial profile, taking for specificity  $E_e =  10^{-2}$ V/$\mu$m. This profile is  similar to the experimental one, shown in Curve a.  Curve e   of  Fig. \ref{Fig02}  shows the calculated spatial intensity profile for an effective increased excitation power of 1 mW. Again, this curve  is similar to both  the experimental  Curve b, taken for an equivalent excitation power, and  Curve d,  implying  that  the calculated profile weakly depends on excitation power.  Finally, Curve f shows the intensity profile calculated using $T_e=60K$. This curve is similar to Curve e showing,  in agreement with the experimental results and with  Eq. \ref{Ee}, that the temperature increase has little effect on the profile. \

\section{ Origin of the large value of the internal electric field}

It is firs shown that the tail in the luminescence spatial  profile cannot be explained by the usual ambipolar model [Eq. \ref{champel1}],  even if a field-activated electron mobility is included. In order to evaluate the electric field in this case, one uses $\nabla n/n  \approx - (\mu_e E  \tau) ^{-1}$, as found from  Eq. \ref{eqdifel} assuming that drift currents are larger than diffusive currents. Further using Einstein's relation, one obtains  

\begin{equation}\label{champel3}
E^2 \exp[-E_e/ |E|)] = \mathscr{F}^2. \frac{1-\frac{\mu_h}{\mu_e} exp[E_e/ |E|)]}{1+\frac{\mu_h}{\mu_e} \exp [E_e/ |E|)]}  \approx  \mathscr{F}^2 
\end{equation} 
	
The electric field  $\mathscr{F}$, given by $\mathscr{F}^{2}=\mathscr{E}/ q \mu_e^* \tau$, is  $10^{-3}$ V/$\mu$m, with the parameter values used in Sec. III, and taking $ \mathscr{E} = 3 meV$.  Numerical resolution of this equation shows that the fraction in the right hand is close to unity and can be approximated as shown in  Eq. \ref{champel3}. Up to  $E_e = 10^{-2}$ V/$\mu$m, Eq. \ref{champel3}  has a solution close to $\mathscr{F}$.  This value is comparable with usual ambipolar fields. It does not interpret the results and is in contradiction with  the starting hypothesis of large drift currents. \

\begin{figure}[tbp]
\includegraphics[clip,width=9 cm] {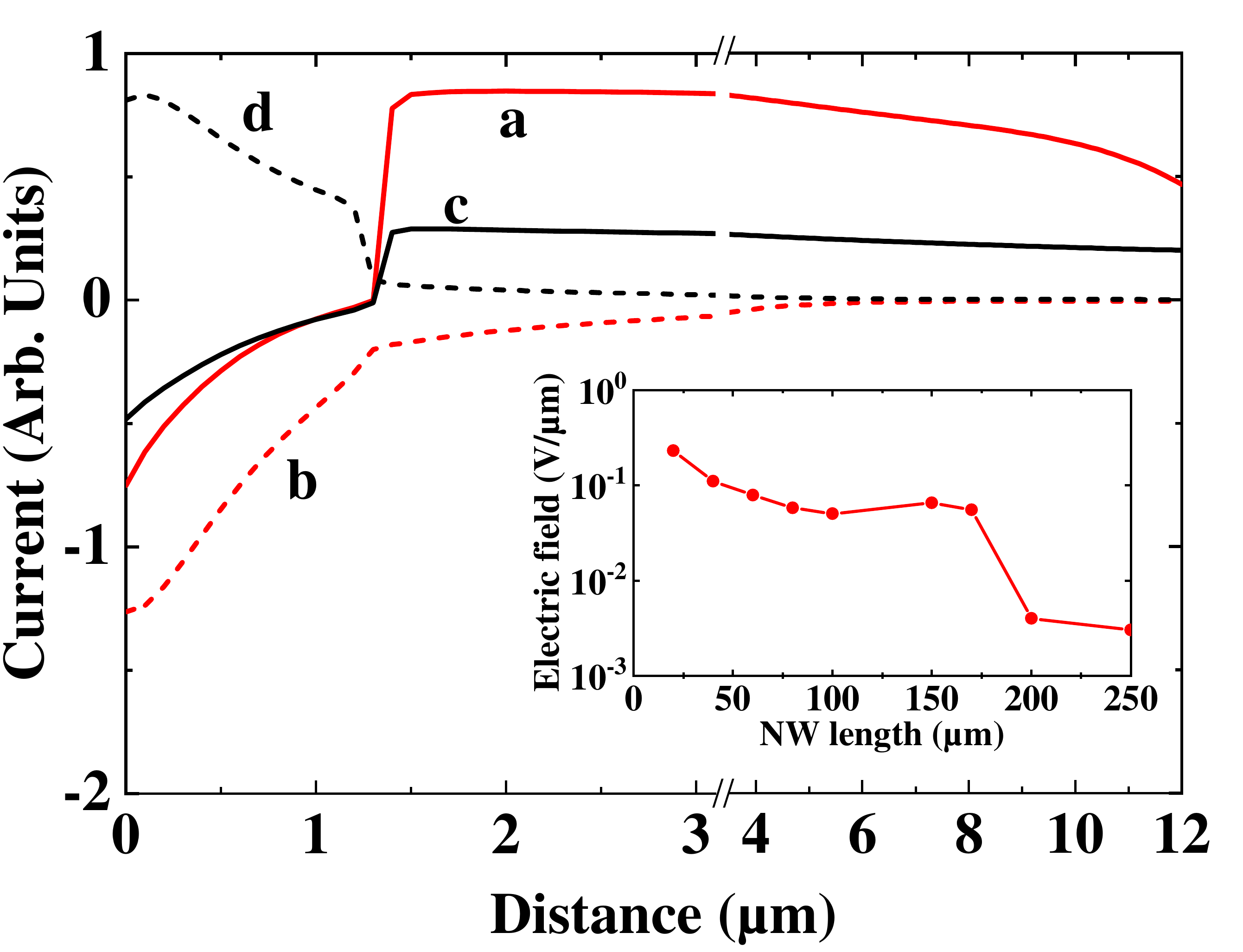}
\caption{Spatial profiles of  electron diffusive (Curve b) and drift (Curve a) currents, as well as of the hole diffusive (Curve d) and drift (Curve c) currents, calculated with $E_e= 10^{-3}$ V/$\mu$m.  The inset shows the electric field value at  10 $\mu$m from the excitation spot, as a function of the NW length. The strong reduction of electric field for a length larger than 150 $\mu$m demonstrates the length-dependent transport. }
\label{Fig04}
\end{figure} 

This failure is not caused by a possible breaking of the hypothesis of charge neutrality, since numerical simulations confirm  its validity  for calculating the electric field \cite{note14}, except in a short stretch near the NW end. We propose that the reason  why Eq. \ref{champel3} cannot explain the large electric field is  that, because of the slow tail, the photocarrier concentrations and currents near  the NW end cannot be neglected. Integration of Eq. \ref{cour} between coordinates $z$ and $z_0$ shows that  it is necessary to  replace Eq. \ref{champel1} by 
\begin{equation}\label{champel2}
	  \vec{\mathscr{J}}(z)  -  \vec{ \mathscr{J}}(z_0) = D_h   \vec{\nabla} p -  D_e \vec{\nabla} n 
\end{equation}  \ 

where $ \vec{\mathscr{J}}(z)  =\vec{E}  \left[ \mu_e (n+n_0) +\mu_h p  \right] $ is the sum of drift  currents at position $z$ in the slow tail.  Here,  $z_0$ is chosen  to be sufficiently large so  that, as shown in Fig. \ref{Fig04}, the diffusive current at $z_0$  is negligible. Inclusion of the negative term $- \mathscr{J}(z_0)$   in Eq. \ref{champel2} should  result in an  increase  of the electric field.  In order to verify this hypothesis, we have calculated the spatial profiles for increasingly large values of the NW length. The  inset of Fig. \ref{Fig04} shows the electric field values at a distance of 10 $\mu$m from the excitation spot. Up to a NW length of  170 $\mu$m, the electric field  weakly depends on distance and is of   $\approx 5 \times 10^{-2}$ V/$\mu$m. For a further increase of NW length, one observes a strong decrease of electric field down to  $3 \times  10^{-3}$ V/$\mu$m that is, to a value comparable with the usual ambipolar regime. It is concluded that, at least for the parameter values chosen here, the anomalous ambipolar transport is amplified by NW finite size effects,  provided the NW length is smaller than   170 $\mu$m. This length is smaller than the maximum length of usual NW's. \ 

The inset of Fig. \ref{Fig04} shows that  the transition between the two regimes occurs over a narrow range of  30 $\mu$m of NW length, while  the electric field is nearly constant before and after the transition. In the same way,  Fig. \ref{Fig03} shows that the transition  as a function of $E_e$ occurs over only a factor of 3 of variation of $E_e$ with profiles nearly independent on $E_e$ before and after the transition.  These features  reveal that the transition between the two regimes occurs through a critical process. This can be understood qualitatively, assuming that $E(z) \approx E(z_0)$ and  that the photocarrier concentrations at $z_0$ are fractions of  their values at position $z$  [$n( z_0)= \xi n(z)$ and $p(z_0) = \xi p(z)$]. The approximate equation  Eq. \ref{champel3} is still valid, provided  $\mathscr{F}$ is divided by $\sqrt{1-\xi}$.  This induces an increase of  $\mathscr{F}$ and therefore of the electric field. This will in turn induce an increase of     $\xi$ since this quantity also depends on electric field, according to $\xi \approx \exp{(z -z_0) /(qE \mu_e \tau) } $]. The quantity $\xi$  is then closer to unity, which will induce a further increase of electric field.\

\section{Conclusion}

it is shown that depleted NW's  on the metallic side of the insulator/metal transition (low $ 10^{17}$ cm$^{-3}$ range) appear as ideal candidates for charge and spin transport : i)  The spatial   profiles  of the luminescence intensity  exhibit a long distance tail, weakly dependent  on  excitation power and temperature, concerning about $10\%$ of the photocarriers and characterized by a decay length  larger than 15 $\mu$m. ii)   The spin polarization  also weakly decreases  with distance. \

A self-consistent  coupled resolution of the drift-diffusion equations, using the  Poisson equation and Boltzmann statistics (Van Roosbroeck model)  shows that the tail occurs because of photocarrier drifting in an internal electric field as large as $10^{-2}$   $V/\mu$m. Two ingredients are crucial for  building up such a large electric field  : i) the dependence of the photoelectron mobility on internal electric field  which strongly increases the elecron mobility and results in a field-assisted transport.  ii) Critical amplification of these effects caused by the NW finite size..\

 Note finally that the above reasoning are only valid for one dimensional systems i. e. if the lateral dimension is smaller than the typical decay length. Indeed, calculations on 2D systems, with increasing $E_e$,   have not revealed any transition to a slow tail.  Since only a limited  number of parameter values was  chosen, these results need to be confirmed by further experimental and theoretical investigations, which are out of the scope of the present work. However, this suggests   that the one dimensional nature of the NW's plays a key role  and that  NW's are better candidates for charge and spin transport than 2D or 3D systems. \

\acknowledgments 

{We are grateful to  V. L. Berkovits and to P. A.  Alekseev for advices in the chemical surface passivation.  This work was supported by Région Auvergne Rhône- Alpes (Pack ambition recherche; Convention 17 011236 01- 61617, CPERMMASYF and LabExIMobS3 (ANR-10- LABX-16-01). It was also funded by the program Investissements d ’avenir of the French ANR agency, by the French governement IDEX-SITE initiative 16-IDEX-0001 (CAP20- 25), the European Commission (Auvergne FEDER Funds). }
 
\bibliographystyle{apsrev}


\begin{thebibliography}{27}
\bibitem[{not({\natexlab{a}})}]{note13}
	\bibinfo{note}{Such result has been verified by introducing a field dependence
		of the hole mobility, according to $\mu_{h}(E)= \mu_{h}^* \exp\left[-\left(
		E_{h}/|E| \right) \right] $ where $ \mu_{h}^*$ is the high-field value of the
		hole mobility, and by performing a numerical calculation of the profile for
		increasing values of $E_{h}$.}
\bibitem[{not({\natexlab{b}})}]{note14}
	\bibinfo{note}{Using Poisson's equation and taking $E=10^{-2}$ $V/\mu$m,
		decreasing with a characteristic distance of $10$ $V/\mu$mwe estimate $ p-n
		\approx 10^{12}$ cm$^{-3}$ that is, 4 orders of magnitude smaller than the
		carrier concentration at the excitation spot.}
\end{thebibliography}

\end{document}